\documentclass[%
aps,
prl,
twocolumn,
superscriptaddress,
longbibliography,
amsmath,amssymb,
floatfix,
nofootinbib, 
]{revtex4-2}

\usepackage{amsmath,amssymb,amsfonts,mathtools}
\usepackage{bm}
\usepackage{physics}
\usepackage{graphicx}
\usepackage{xcolor}

\usepackage[normalem]{ulem}

\usepackage{hyperref}

\newcommand{\lag}{\mathcal{L}}
\newcommand{\der}{\mathrm{d}}

\newcommand{\rstar}{R_\text{M}}

\begin{document}

\title{An empirically constrained low density regime Covariant Modified Gravity: \\
accurate rendering of galactic rotation curves and lensing resulting in present accelerated cosmological expansion.}

\author{Hernandez, X.}
\email{xavier@astro.unam.mx}
\affiliation{Instituto de Astronom\'ia, Universidad Nacional Aut\'onoma de 
M\'exico, AP 70-264, Ciudad de M\'exico 04510, M\'exico}

\author{Sergio Mendoza}
\email{sergio@astro.unam.mx}
\affiliation{Instituto de Astronom\'ia, Universidad Nacional Aut\'onoma de 
M\'exico, AP 70-264, Ciudad de M\'exico 04510, M\'exico}

\author{Sarah\'{i} Silva}
\email{sgarcia@astro.unam.mx}
\affiliation{Instituto de Astronom\'ia, Universidad Nacional Aut\'onoma de 
  M\'exico, AP 70-264, Ciudad de M\'exico 04510, M\'exico}

\author{Edgar Gasper\'in}
\email{e.gasperin@nucleares.unam.mx}
\affiliation{Instituto de Ciencias Nucleares, Universidad Nacional Autónoma de México, Apartado Postal 70-543, Ciudad de México 04510,
Mexico}

\date{\today}

\begin{abstract}

 We present a covariant modified gravity theory calibrated to match all the galactic phenomenology
 usually attributed to dark matter in the low acceleration regime, including the flat rotation curves
 of massive tracers, and the deflection angles of photons as inferred through gravitational lensing. 
 We first derive a fully covariant description of galactic space-times under a spherically
 symmetric and static approximation. This description accounts for flat rotation curves,
 observed gravitational lensing phenomenology and the Tully-Fisher mass scalings of the above.
The resulting covariant description is then used to constrain the parameters of a minimally coupled
gravitational action using power laws for both the Ricci scalar and the matter Lagrangian. The action
presented hence leads to field equations whose trace, has as solutions metric coefficients which 
reproduce flat rotation curves and gravitational lensing observations, without including any dark matter.
Using the same action for a FLRW metric results in a cosmology where a dust component behaves as effective
dark energy with an equation of state $\omega=-5/6$, without the explicit inclusion of any dark energy component.
 
\end{abstract}

\maketitle

\section{Introduction}
Discerning between modified gravity theories and the presence of dark matter
remains one of the most persistent challenges in
modern astrophysics and cosmology. While the standard $\Lambda$CDM paradigm
can successfully model a wide range of observations---from the cosmic
microwave background to large-scale structure---it requires that
approximately 85\% of the matter content of the Universe be in the form of
an as-yet-undetected hypothetical non-baryonic component. At galactic scales, the
observed flatness of rotation curves and the tight correlations between
baryonic mass and rotational velocity (the Tully--Fisher relation) have
motivated numerous alternative explanations, including modifications of
Newtonian dynamics (MOND) \cite{Milgrom1983}.

A first crucial point is that the gravitational anomalies attributed to the presence of a hypothetical and
dominant dark matter component under GR, appear only at acceleration scales lower than a critical value
identified by MOND of $a_{0} = 1.2 \times 10^{-10}$ms$^{-2}$. At higher accelerations no dark matter
is required by astronomical constraints, and hence no detectable modifications to GR are allowed by the data.

MOND is empirically grounded upon the observation of flat rotation curves,
and in the low velocity regime, has been remarkably successfully in
reproducing rotation curve observations across an impressive range of galactic
masses and types, e.g. \cite{McGaugh2000}, \cite{Lelli2017}. Being a Newtonian framework,
MOND is incapable of addressing relativistic phenomena such as gravitational lensing
and cosmology. Attempts to extend MOND into the relativistic regime are faced with the
problem of guessing an underlying covariant theory of which only the low velocity
limit is known. Such attempts have so far failed at a reaching a consistent description
of relativistic phenomena. This can be seen for example in \cite{Sagi2010} showing TeVeS \cite{Bek2004} yields
superluminal gravitational waves, or \cite{Mistele2023} showing the AeST theory
of \cite{Skordis2021} is in conflict with observations of gravitational lensing at galactic scales.

Indeed, even at Newtonian level, in moving away from the empirical galactic regime
upon which MOND is grounded, in going to galaxy clusters scales, a significant
mass deficit appears in MOND e.g. \cite{Benoit2025}, \cite{Hernandez2026}.

In going to covariant extended theories of gravity e.g. \cite{Faraoni2010,Capozziello2011},
$f(R)$ theories and their generalisations with matter
couplings have received particular attention due to their ability to
explain both cosmic acceleration and galactic dynamics without exotic dark
components. However, no such attempt has simultaneously reproduced the
MOND success at describing flat rotation curves, the Tully-Fisher relation,
gravitational lensing observations and the present-day accelerated cosmic expansion.



Here we take the approach of grounding a covariant theoretical development
directly upon both the low velocity phenomena of galactic rotation curves and
their observed mass scalings, together with observed relativistic gravitational
lensing phenomenology and corresponding mass scalings. To this end we start by
deriving the full covariant structure of galactic space-times and use it to
calibrate a gravitational action. The resulting minimal theory naturally has a
MOND-like low velocity limit, satisfies gravitational lensing observations, and
can also reproduce galaxy cluster dynamics, as well as yielding gravitational
wave propagation speeds rigorously equal to c and a late time accelerated
cosmological solution, consistent with recent observational constraints.


The paper is organised as follows: the following section derives a fully
covariant description of galactic space-times in the low acceleration regime. Next
we present the theoretical framework, including the action, field equations,
and trace equation for dust. The last sections calibrate the parameters of the
theoretical model through the empirical constraints first presented, develop the
cosmological consequences of the framework presented and give a final discussion.


\section{Low acceleration Empirical Galactic Spacetimes}
\label{sec:empirics}

We begin with a general spherically-symmetric and static metric as a first approximation to galactic
space-times in the low acceleration regime:

\begin{equation}\label{Metric_SSS}
  \der s^2=g_{00}(r)c^2\der t^2+g_{11}(r)\der r^2+r^2\der\Omega^2.
\end{equation}  

The first empirical constraint to be introduced is that at the low velocity limit, equilibrium rotation velocities
of test particles on circular orbits are flat, this restriction yields:

\begin{equation}
g_{00}(r)=-1-2\beta^{2}\ln(r/\rstar),
\label{empiric-00}
\end{equation}

\noindent working under a \( (-,+,+,+) \) metric signature used throughout. 
In the above we recognise the logarithmic Newtonian potential of a galactic flat rotation curve of amplitude
$V=\beta c$, hence, $\beta$ is the flat rotation velocity in units of the speed of light. $\rstar$ is a scale radius introduced
for dimensional reasons, while it has no bearing on observables as it is just an additive constant on the
potential which vanishes when deriving to obtain forces and velocities it can be adjusted e.g. to ensure continuity in $g_{00}$
at an inner transition to a Schwarzschild region.

The second empirical constraint to consider is the well established gravitational lensing phenomenology of constant
deflection angles $\Delta \varphi$ given by (see e.g. \cite{Mendoza2013}):

\begin{equation}
\Delta \varphi=\pi \beta^{2}.
\end{equation}

This relation can then be introduced as a constraint into:

\begin{equation}
\Delta \varphi (r_{0}) = 2\int_{r0}^{\infty} \frac{\left[-g_{00}(r) g_{11}(r)
     \right]^{1/2}}{r\left[(r/r_{0})^{2}g_{00}(r_{0})-g_{00}(r) \right]^{1/2}} dr   -\pi,
\end{equation}

\noindent the integral giving the deflection angle as a function of the impact
parameter, $r_{0}$, and the metric coefficients $g_{00}$ and $g_{11}$. Taking also $g_{00}$ form the flat rotation curve
restriction given above, (see e.g. \cite{Mendoza2013} or \cite{Campigotto2017}) allows to solve for $g_{11}(r)$ as:

\begin{equation}
g_{11}(r) =1+2\beta^{2}.  
\label{empiric-rr}
\end{equation}

Equivalently, the GR inference of singular isothermal halos for both rotation curve and lensing observations of
galaxies leads to the same metric coefficients as derived above for a zero pressure singular isothermal matter distribution
in the low velocity limit through equation (25.82) in \cite{Misner1973},
considering the approximation $A^{x} \approx 1+x\ln(A)$,
which is accurate as for all observed galaxies, as $\beta <<1$.

Eqs. (2) and (5) now give a relativistic description of galactic space-times in the spherical approximation, although
still particular to the choice of metric introduced in eq.(1). To reach a fully covariant description we can calculate the
curvature scalars corresponding to the metric coefficients obtained. Given the high degree of symmetry of the spherically
symmetric and static ansatz introduced, the Chern-Pontryagin scalar is zero, so that the only non-zero curvature
scalars up to second order are:

\begin{equation}
  R= 2 \left(\frac{\beta}{r}\right)^{2}, S=3\left( \frac{\beta}{r}\right)^{4}, I_{1}=\frac{64}{3}\left( \frac{\beta}{r}\right)^{4};
  K=28\left( \frac{\beta}{r}\right)^{4},
\label{empR}
\end{equation}

\noindent the Ricci scalar, the contraction of the trace-free Ricci tensor and
the full contraction of the Weyl tensor, respectively, see
e.g. \cite{Cherubini2002}, \cite{Hernandez2019}. The Kretschmann scalar, $K$, is not
independent of the preceding quantities, but is given for completion, on account of it being a familiar quantity.
Notice that the covariant description of galactic space-times given is entirely general to
any geometrodynamical theory, GR or otherwise, and is simply an inescapable reflection of empirical observed galactic dynamics.
Of course, under GR, this space-time implies the presence of a dominant close to singular isothermal dark matter halo.

Note that the above results are given only to leading order in $\beta$. With the largest
known galactic systems having values of $\beta<10^{-3}$, and only even
powers of $\beta$ appearing in the full relations, the above expressions
suffice for a first covariant description of galactic space-times. Further,
the empirical constraints introduced in this section are also not exact
in a mathematical sense, but only suitable representations of astronomical
data subject to observational errors.

Figure 1 shows a comparison of the radial scalings of $R^{2}$ and $K$ for the Schwarzschild solution and the empirical
galactic space-times. We see that the galactic space-time gives a Kretschmann scalar
which falls less abruptly than in the Schwarzschild case, but only presents a change of slope at the transition radius.
Indeed, notice that when written in terms of the Keplerian velocity about a point mass, $\beta_{Sw}$, the Kretschmann
scalar is $K_{Sw}=48(\beta_{Sw}/r)^{4}$, appearing very analogous to the galactic quantity. The
Ricci scalar on the other hand has a markedly different behaviour, switching from a rather natural scaling with the square root
of $K$ in the galactic regime, to being zero in the Schwarzschild one.
This difference makes both space-times fundamentally different. In the figure the transition is shown as abrupt, merely for illustrative
purposes, it is hard to establish observationally the details of this transition in detail, and for this amongst other reasons, in what
follows we shall be concerned exclusively with the well-established galactic space-times far from any inner transition to GR.

\begin{figure}
\centering
\includegraphics[width=0.5\textwidth, height=210pt]{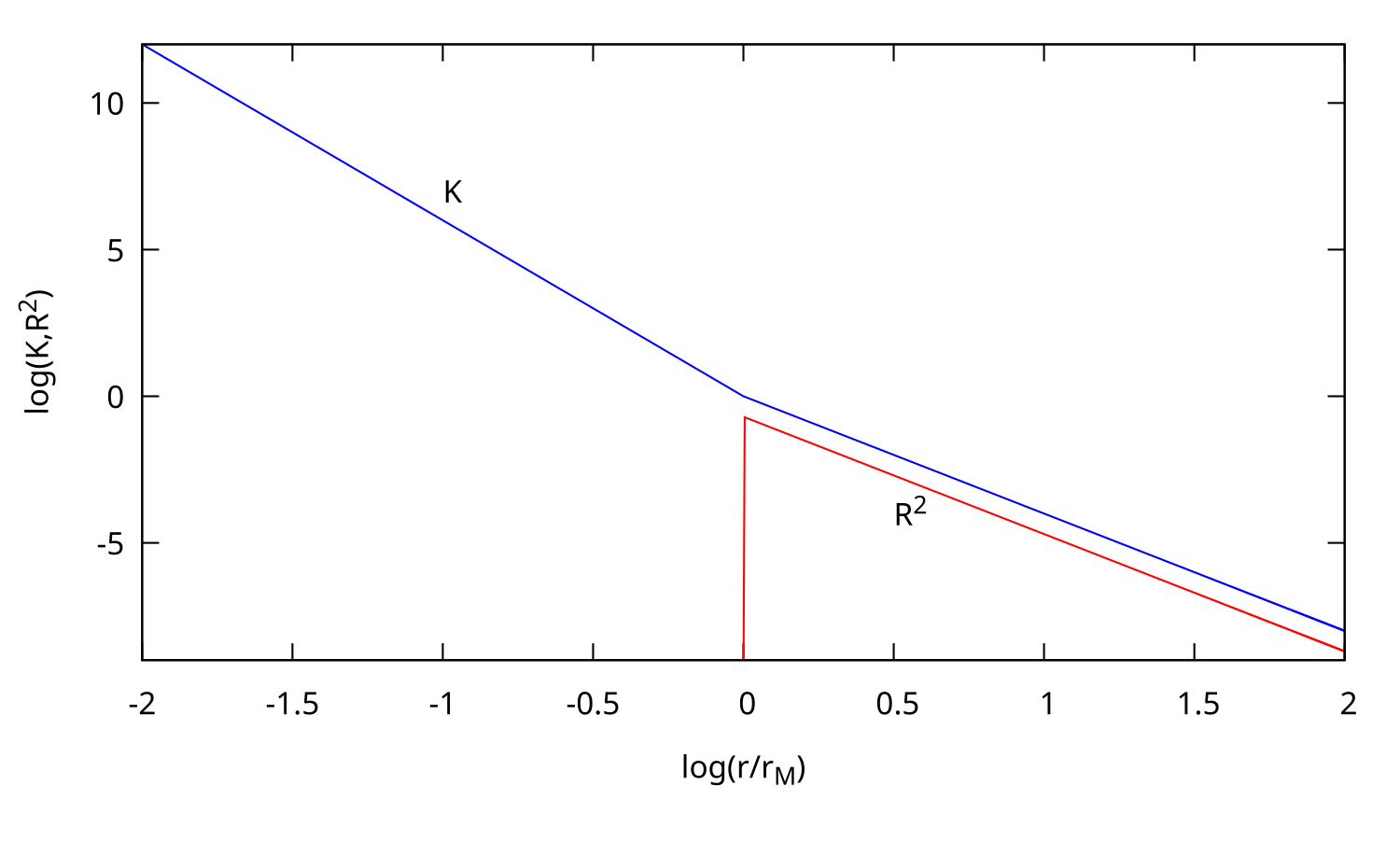}
\caption{Comparison of the radial scaling of the Ricci and Kretschmann scalars in the Schwarzschild solution at small
distances and the empirical galactic space-time at large distances.}
\end{figure}

A final empirical constraint comes from the tight scaling observed between $\beta$ and the baryonic mass of
a galaxy, which we shall identify with the Schwarzschild mass of the system, $M$. This is referred to as the
baryonic Tully-Fisher (henceforth BTF) relation, where the amplitude of a galaxies' rotation curve is correlated to
its total baryonic mass through a power law scaling:

\begin{equation}
  \beta c=(G a_{0})^{1/4} M^{1/4},
\label{TFobs}
\end{equation}

\noindent as first presented by \cite{McGaugh2000}, where $M$ is the baryonic mass of a galaxy and $a_{0}$ the MOND acceleration
constant. Notice that determining such a scaling
requires accurate and independent measurements of both flat rotation curve velocities and
total baryonic mass inferences. These last are constructed through gas and star fractions, and crucially,
the assumption or inference of mass to light ratios, $M/L$, for the galaxies studied to obtain masses from the observable
luminosities. If one assumes that $M/L$ values for the disks of spiral galaxies are all constant, then
it is obtained that the BTF index is, within observational uncertainties, $1/4$ as given above, \cite{Lelli2016}.
It is interesting that this result matches original predictions of MOND predating it by more than 15 years. However,
a number of astronomical trends imply more massive disks are more gas rich and contain older and more metal rich populations,
all of which operate in the same direction of leading to systematically higher $M/L$ ratios as one moves from low mass
galactic disks to high mass galactic disks. Indeed working without assuming a constant $M/L$ ratio for galactic disks,
but using values appropriate at each galaxy to the gas and stellar populations actually present, leads to BTF indices
higher than $1/4$. For example, \cite{Ravi2026} recently used a high quality sample of 5743 galaxies from the MANGA sample
to derive a BTF inverse index of $n=3.86^{+0.92}_{-0.62}$. While certainly consistent with $1/4$, the inference centres
upon slightly higher values. It is clear that this final crucial observational constraint is still uncertain in the detail.
We shall take the central simple value of $1/4$ given above, but note that this value could change once the observational
scenario is better constrained.

Thus, eqs (2), (5) (6) and (7) form the full description of galactic space-times to be used as constraints for the development
of a covariant modified gravity scenario in the low acceleration regime, by construction, consistent with all of the galactic
phenomenology attributed to a hypothetical dominant dark matter component when modelling rotation curves and lensing observations
under GR. Clearly, in the high acceleration regime where no gravitational anomalies are evident with respect to GR, standard
gravity must apply.

\section{Action and Field Equations}
\label{sec:framework}

For the low acceleration regime described geometrically in the previous section, we consider a generalised action:

\begin{equation}
S=\frac{1}{2\kappa^2}\int\der^4x\sqrt{-g}\,F(R)
+\int\der^4x\sqrt{-g}\,H(\lag_ {\mathcal{M}}),
\label{action}
\end{equation}

\noindent where $\kappa$ is a coupling constant (given by $\kappa^2=8\pi
G / c^4 $ in the case of general relativity),  $\lag_ {\mathcal{M}}$ is the matter Lagrangian density
and,

\begin{equation}
F(R)=R^{b}, \qquad  H(\lag_ {\mathcal{M}})=\lag_ {\mathcal{M}}^m,
\label{t01}
\end{equation}

\noindent with $b$ and $m$ indices to be determined. 
For a perfect fluid 
$\lag_ {\mathcal{M}}=e= \rho c^2 + \Pi$, where \( e  \) is the total energy density,  \( \rho \) is the rest mass energy density and  $\Pi$ is the pure internal energy density of the fluid, which for dust reduces to
$\lag_ {\mathcal{M}}=\rho c^2$.  This choice ensures that the
matter Lagrangian is well-defined and avoids the ambiguities associated
with the pressure prescription
\cite{Harko2011,sarahi01,sarahi02,sarahi03}.

The energy-momentum tensor is defined as

\begin{equation}
T_{\mu\nu}=-\frac{2}{\sqrt{-g}}\frac{\delta(\sqrt{-g}\lag_ {\mathcal{M}})}{\delta g^{\mu\nu}},
\end{equation}

\noindent with trace $T=g^{\mu\nu}T_{\mu\nu}$.

The null variations of the action \eqref{action} with respect to the metric
$g^{\mu\nu}$ yield the following field equations:

\begin{multline}
F_R R_{\mu\nu}-\frac{1}{2} F g_{\mu\nu}
+(g_{\mu\nu}\Delta-\nabla_\mu\nabla_\nu)F_R \\
= \kappa^2 H_{\lag_ {\mathcal{M}}} (T_{\mu\nu}-\mathcal{L}_\mathcal{M} g_{\mu\nu} )  + \kappa^2 H g_{\mu\nu},
\end{multline}
\noindent where $F_R:=\partial F/\partial R=bR^{b-1}$,
$H_{\lag_ {\mathcal{M}}}:=\partial H/\partial\lag_ {\mathcal{M}} = m \lag_ {\mathcal{M}}^{m-1} $, and the Laplace-Beltrami operator is given by $\Delta:= g^{\mu\nu}\nabla_\mu\nabla_\nu$.

For dust, the pressure $p=0$ and thus \( \Pi = 0 \) and so the trace of the energy-momentum tensor is
$T= -\mathcal{L}_\mathcal{M} = -e = - \rho c^2$. 

Taking the trace of the field
equations, we obtain

\begin{equation}
F_R R-2F +3\Delta F_R = \kappa^2 H_{\mathcal L_{\mathcal{M}}} \left( T - 4 \mathcal{L_\mathcal{M}} \right) + 4 \kappa^2 H, 
\end{equation}

\noindent Direct substitution of equation~\eqref{t01} into the previous relation for the case of dust, we get:

\begin{equation}
(b-2)R^{b}+3b\,\Delta R^{b-1}= \eta T ^m  = \eta \left( \rho c^2 \right)^m, 
\label{traceequation}
\end{equation}

\noindent with $\eta := \kappa^{2} (4- 5 m)$.

\section{Weak-Field Limit and Fixing $b$, $m$ and $\kappa$ Through Empirical Constraints}
\label{sec:weakfield}


In what follows we will use standard Parametrised Post Newtonian  (PPN) notation as presented in \cite{will}.  As such, when perturbing quantities we
will use the notation \( \mathcal{O}(n) \) meaning that that quantity is of order \( \mathcal{O}(1/c^n) \), which is to say the particular quantity is proportional to \( 1/c^n \).

For the static, spherical symmetric metric at 
\( \mathcal{O}(2)  \) of approximation \footnote{Note that in general terms, the metric can be approximated at \( \mathcal{O}(2) \) in conformal coordinates as:
\begin{equation}
   \mathrm{d}s^2 = -(1 +  \frac{2 \phi}{c^2} ) c^2 \mathrm{d}t^2 + (1 +  \frac{2 \psi}{c^2}  ) \mathrm{d}x^k \mathrm{d}x_k,
\end{equation}

\noindent but since the empirical metric requires the values given in equation~\eqref{empiric-00} and~\eqref{empiric-rr}, then as shown by \cite{mendoza2023metric}, it follows that \( \phi = \psi\) and so, transforming this conformal metric to spherical coordinates, equation~\eqref{perturbedmetric} is obtained according to the results in~\cite{mendoza2023metric}.
}: 

\begin{gather}
g_{00}=-1+g_{00}^{(2)}+\mathcal{O}(4) = -1 - \frac{ 2 \phi }{c^2}, 
  \nonumber \\ 
g_{11}=1+g_{11}^{(2)}+\mathcal{O}(4) = 1 + \frac{ 2 r \left( \mathrm{d} \phi / \mathrm{d} r \right) }{ c^2 } ,
\label{perturbedmetric}
\end{gather}

\noindent where the superscript $(2)$ denotes terms of order $\mathcal{O}(2)$. The Ricci scalar reduces to
\begin{equation}
R^{(2)} \simeq \frac{2}{c^2}\nabla^2\phi,
\end{equation}
where $\nabla^2$ is the Laplacian operator. For a test particle in a
circular orbit with velocity $v$, we have $\nabla^2\phi\sim v^2/r^2$.

In vacuum, \( \rho = 0 \) and the first term on the left-hand side of equation~\eqref{traceequation}, at the lowest order of \( R \), i.e. \( R^{(2)} \),  of  is of order \( \mathcal{O}(2b) \) and the second is of order \( \mathcal{O}(2b-2)\).  In other words, for \( b > 1 \) the leading term is the second one and thus, the trace equation in vacuum at the lowest perturbation order is given by \cite{Mendoza2013}:

\begin{equation}
\Delta\left(R^{(2)}\right)^{b-1} = \frac{2}{c^2} \nabla^2 \left( \nabla^2 \phi \right)^{b-1} = 0., 
\label{vacuum-trace}
\end{equation}

The spherically symmetric solution of this equation is
\begin{equation}
\left(R^{(2)}\right)^{b-1}=\frac{A}{r},
\end{equation}
where $A$ is an integration constant \cite{Mendoza2013}. Requiring flat
rotation curves and consistency with empirical lensing observations
through \( R = 2 \left( \beta / r \right)^2 \) of equation~\eqref{empR}
forces 

\begin{equation}
\frac{1}{b-1}=2\quad\text{i.e.}\quad b=\frac{3}{2}.
\end{equation}

This uniquely fixes the curvature exponent. The choice $b=3/2$ is therefore
imposed by the empirical galactic space-time constrains presented previously.

With $b=3/2$, the vacuum solution of the trace of the field equations becomes:

\begin{equation}
R^{(2)}=\frac{A^2}{r^2} 
\end{equation}
where $A^2=2\beta^{2}$ is a dimensionless quantity of $\mathcal{O}(2)$.


We can now confirm the validity of the $b=3/2$ solution found by recovering the
empirical metric coefficients directly as a solution to the field ecuations found.

Using the field equations in vacuum at order $\mathcal{O}(2)$, we obtain
the following system of coupled differential equations:
\begin{align}
r^2 g_{00,rr}^{(2)}+3r g_{00,r}^{(2)}+2A^{2} &= 0, \\
r g_{11,r}^{(2)}+g_{11}^{(2)}+\frac{k_1}{r^2}-A^{2} &= 0,
\end{align}
with solutions \cite{Mendoza2013}
\begin{align}
g_{00}^{(2)}(r) &= -A^{2}\ln\left(\frac{r}{\rstar}\right)
+\frac{k_1}{r^2}, \\
g_{11}^{(2)}(r) &= \frac{k_1}{r^2}+\frac{k_2}{r}+A^{2}.
\end{align}
The constant $k_1$ is fixed by requiring flat rotation curves: $k_1=0$.
Furthermore, matching the empirical $g_{11}$ coefficient found previously through
requiring consistency with gravitational lensing observations
requires $k_2=0$ e.g. \cite{Mendoza2013}. Thus, we obtain
\begin{align}
g_{00}(r) &= -1-A^{2}\ln\left(\frac{r}{\rstar}\right), \\
g_{11}(r) &= 1+A^{2}.
\end{align}
These results are exact in the weak-field limit and provide the explicit
form of the metric coefficients. Comparison to the empirical metric coefficients
for galactic space-times presented previously shows agreement.

%

\subsection{Determination of the Coupling Constant}
\label{sec:coupling}

As mentioned above, the leading order of the left-hand side of trace equation~\eqref{traceequation} is the second term, which is of $ \mathcal{O}(2b-2) $.   If the right-hand side of the trace equation is of order \( \mathcal{O}(q) = \mathcal{O}(\eta) - 2 m\), with \( q < 2b -2 \) then the right-hand side would be the leading term of the whole trace equation and thus \( \rho  = 0 \)  at that perturbation order.  Also, if \( q > 2b -2 \), then the matter term does not appear in the perturbed trace and the vacuum trace equation~\eqref{vacuum-trace} is valid.  However, if \( q = 2b - 2\), then the field equation is given by:

\begin{equation}
\Delta\left(R^{(2)}\right)^{b-1} = \nabla^2 \left( \nabla^2 \phi \right)^{b-1} = \alpha T^m, 
\label{vacuum-full}
\end{equation}

\noindent with \( \alpha := \eta / 2b = \kappa^2 (4-5m) / 3b \).

Under this last assumption, which is the only relevant option yielding a curvature-matter coupling, let us  consider a point mass source \( M \) 
producing the gravitational field. To fix the coupling constant
$\kappa$, we start by integrating the previous trace equation over a small ball of radius $\epsilon$:
\begin{equation}
\int_{V_\epsilon}\nabla^2 ( R^{(2)} )^{1/2}\,\der V
=\alpha\int_{V_\epsilon} T^m\,\der V.
\end{equation}
Using the fact that $\nabla^2(1/r)=4\pi\delta^{(3)}(\mathbf{r})$, the left-hand side gives
$4\pi 2^{1/2} \beta$. The right-hand side requires the
renormalization of the composite operator $\rho^m$, and so
\begin{equation}
\int_{V_\epsilon}\rho^m\,\der V:=M^m l_{c}^{3(1-m)}.
\end{equation}
\noindent where $M$ is an effective mass.
This is the standard prescription for handling the singular point-mass
source \cite{Collins1984}, where $l_{c}$ is a quantity with units of length introduced for dimensional
consistency. Since the only such quantity allowed by the relevant physical constants present ($G, c$ and $a_{0}$)
is a multiple of $c^{2}/a_{0}$, we shall take $l_{c}=\gamma(c^{2}/a_{0})$, with $\gamma$ a dimensionless constant
expected to be of order unity. Therefore,
\begin{equation}
4 \sqrt{2} \, \pi \beta = \alpha \gamma^{3(1-m)} (c^{2}/a_{0})^{3(1-m)} M^{m}. 
\end{equation}

Using the Tully-Fisher scaling to substitute $\beta$ for $M$ on the left hand side now yields:
\begin{equation}
4 \pi c^{-1} M^{1/4} \, \sqrt{2 G a_{0}} =\alpha \gamma^{3(1-m)} (c^{2}/a_{0})^{3(1-m)} M^{m}, 
\end{equation}

%
%

\noindent from which $m=1/4$ and $\alpha \gamma^{9/4}= \pi (32)^{1/2} G^{1/4} a_{0}^{5/2}c^{-11/2}$.

%

This determines the constant $\alpha$ in the trace of the field equation in terms of observable quantities and the dimensionless
constant $\gamma$. The index on the matter side of the action is hence fixed, is indeed given directly, by the index of
the baryonic Tully-Fisher relation. 
In order to determine this remaining free constant $\gamma$, we turn to cosmology in the following section.

Note that since \( m = 1/4 \) and so \( \alpha \propto c^{-11/2} \) , then the right-hand side of equation~\eqref{vacuum-full}  is of \( \mathcal{O}(5) \).  In other words, in the weak field limit of approximation, the right-hand side of the trace equation is negligible, and hence our spherical symmetric and static solution is only relevant as a vacuum solution.

While working at the Newtonian level in MOND, forcing a flat rotation curve fixes the
predicted index of the Tully-Fisher scaling at $1/4$, at this more general covariant level we see that
fixing flat rotation curves and lensing fixes the index of the geometric power law, $b=3/2$,
while the predicted index of the Tully-Fisher relation, remains an independent parameter determined by
the matter Lagrangian power index $m$.

This feature is interesting because
a value of the Tully-Fisher index slightly above $1/4$ implies a steeper dependence of dynamical velocity as a function of mass,
and hence the option of fitting galaxy cluster dynamics without the need for any additional mass. This last, as
the consistency with galactic rotation curves and lensing phenomenology, in the absence of any dark matter.

Modified gravity $f(R)=R^{3/2}$ models have been explored before, starting with \cite{Capozziello2009} who showed that for this model
galactic flat rotation curves result, and that enhanced lensing above GR expectations would also be present. Then, \cite{Mendoza2013}
showed that the resulting lensing is indeed compatible with observations. However, both of the above previous studies kept the standard
linear dependence of the right hand side of the action on the energy-momentum tensor. As a result, fitting the BTF scaling in
\cite{Capozziello2009} was done through forcing an {\it ad hoc} explicit mass dependence of the parameters of the solution to the
field equations, while in \cite{Mendoza2013}, fitting this observed mass-velocity scaling was introduced through forcing an
{\it ad hoc} explicit mass dependence on the coupling constant of the theory. Compared to these somewhat contrived previous studies,
the developments presented here have the advantage of recovering the observed galactic mass-velocity scaling directly at the level of
the action through the power law dependence of the matter-energy tensor introduced, with the coupling constant being indeed a constant.
Further, as will be shown in the following section, this development naturally results in an accelerated cosmological expansion compatible
with recent observations.

\section{Cosmology}
\label{sec:cosmology}

In this section we take the trace of the field equations developed previously and apply them in the cosmological setting
of a FLRW metric. In this case, the Ricci scalar is now given by:

\begin{equation}
R=\frac{6}{c^{2}}\left[ \frac{\ddot{a}}{a} + \left( \frac{\dot{a}}{a} \right)^{2} \right]\,\,\,= \frac{6\dot{a}}{(c a)^{2}}
                 \frac{d(\dot{a}a)}{da},
\end{equation}

\noindent where $a$ is the cosmological scale factor. The trace of the field equation now becomes:

\begin{equation}
\frac{-1}{2} R^{3/2} + \frac{1}{c^{2}} \frac{d^{2}(R^{1/2})}{dt^{2}} + \frac{3 H}{c^{2}} \frac{d(R^{1/2})}{dt} =- \alpha \rho^{1/4}.
\end{equation}

\noindent Here and in what follows \( H := \dot{a} / a \) is the Hubble parameter. 

Introducing a dust component $\rho=(3 \Omega_{m} H_{0}^{2}/8 \pi G a^{3})$, and keeping temporarily only the first term on the L.H.S.
yields:
\vskip 0pt

\begin{equation}
\left( \frac{1}{2} \left[ \frac{6\dot{a}}{(c a)^{2}} \frac{d(\dot{a}a)}{da} \right] \right)^{3/2} = \left( \frac{(32)^{1/2}\pi G^{1/4} a_{0}^{5/2}}
     {\gamma^{9/4} c^{11/2}}\right) \left(\frac{3 \Omega_{m} H_{0}^{2}}{8 \pi G a^{3}} \right)^{1/4},
\end{equation}

\noindent and hence:

\begin{equation}
\frac{6\dot{a}}{(c a)^{2}} \frac{d(\dot{a}a)}{da} = \left( \frac{(128)^{1/3}\pi^{2/3} a_{0}^{5/3}}
     {\gamma^{3/2} c^{11/3}}\right) \left(\frac{3 \Omega_{m} H_{0}^{2}}{8 \pi a^{3}} \right)^{1/6},
\end{equation}

\noindent which can be integrated to yield:

\begin{equation}
H^{2}=\frac{(8 \pi)^{1/2}(16)^{1/3}}{7 \gamma^{3/2}} \left( \frac{a_{0}}{3c} \right)^{5/3} H_{0}^{1/3} \frac{\Omega_{m}}{a^{1/2}}.
\end{equation}

The solution to the above equation is simply $a\propto t^{4}$, which leads to the Ricci scalar being $R=168/(ct)^{2}$, and in turn to
the first term in the trace equation of eq.(34) scaling with $(ct)^{-3}$. It is trivial to check that for this $a\propto t^{4}$ solution,
the two terms which were dropped from eq(34) also have the same $(ct)^{-3}$ scaling, at about 10\% of the value of the first term.
Thus, we see that the cosmological solution presented is in fact exact for the full trace of the field equations in the FLRW case, after
a small scaling of the coefficient of the first term in eq(34) by $(94/84)$ as used to obtain the full solution presented below.

We can now compare the modified Friedmann equation of eq.(37) to that of a standard GR case for a single component:

\begin{equation}
H^{2}=H_{0}^{2}\left( \frac{\Omega}{a^{3(1+\omega)}} \right).
\end{equation}

\noindent We see that if a reality described by the modified Friedmann equation is forced into a comparison with a standard Friedmann
equation, the single dust component introduced, which explicitly dilutes with $a^{-3}$, will be seen as a dark energy component with
an equation of state given by $\omega=-5/6$. This parameter, slightly above a classical cosmological constant, is in fact consistent
with recent direct determinations from DESI of broadly $-0.9<\omega<-0.7$. Notice that while the relativistic developments of the
first sections as applied to galactic space-times are only accurate to leading order deviations from Minkowski spacetime, all the
cosmological developments of this section are exact solutions of the proposed low density regime action.

Further, we see that evaluating eq.(37) today at $a=1$ and $H=H_{0}$, imposes a scaling of $a_{0} \approx c H_{0}$. Just as in standard
gravity, the mass-velocity scaling of the experiment used to calibrate the theory ($G$ through Kepler's laws, $a_{0}$ through the baryonic
Tully-Fisher scaling) enters as part of the coupling constant in the relativistic extension of the theory, and then sets the scaling of
the cosmological parameters under a FLRW model.

Finally, taking observational values for $a_{0}=1.2 \times 10^{-10}$ ms$^{-2}$, $H_{0}=70$ km s$^{-1}$ Mpc$^{-1}$ and $\Omega_{m}=0.05$ allows
to finally evaluate $\gamma=0.3$, which allows to fully fix the coupling constant of the low density theory proposed. At this point
we have taken a universe filled only with the only presently relevant component of whose existence we are certain, so that $\Omega_{m}$
reflects only baryons.

\begin{figure}
\centering
\includegraphics[width=0.5\textwidth, height=210pt]{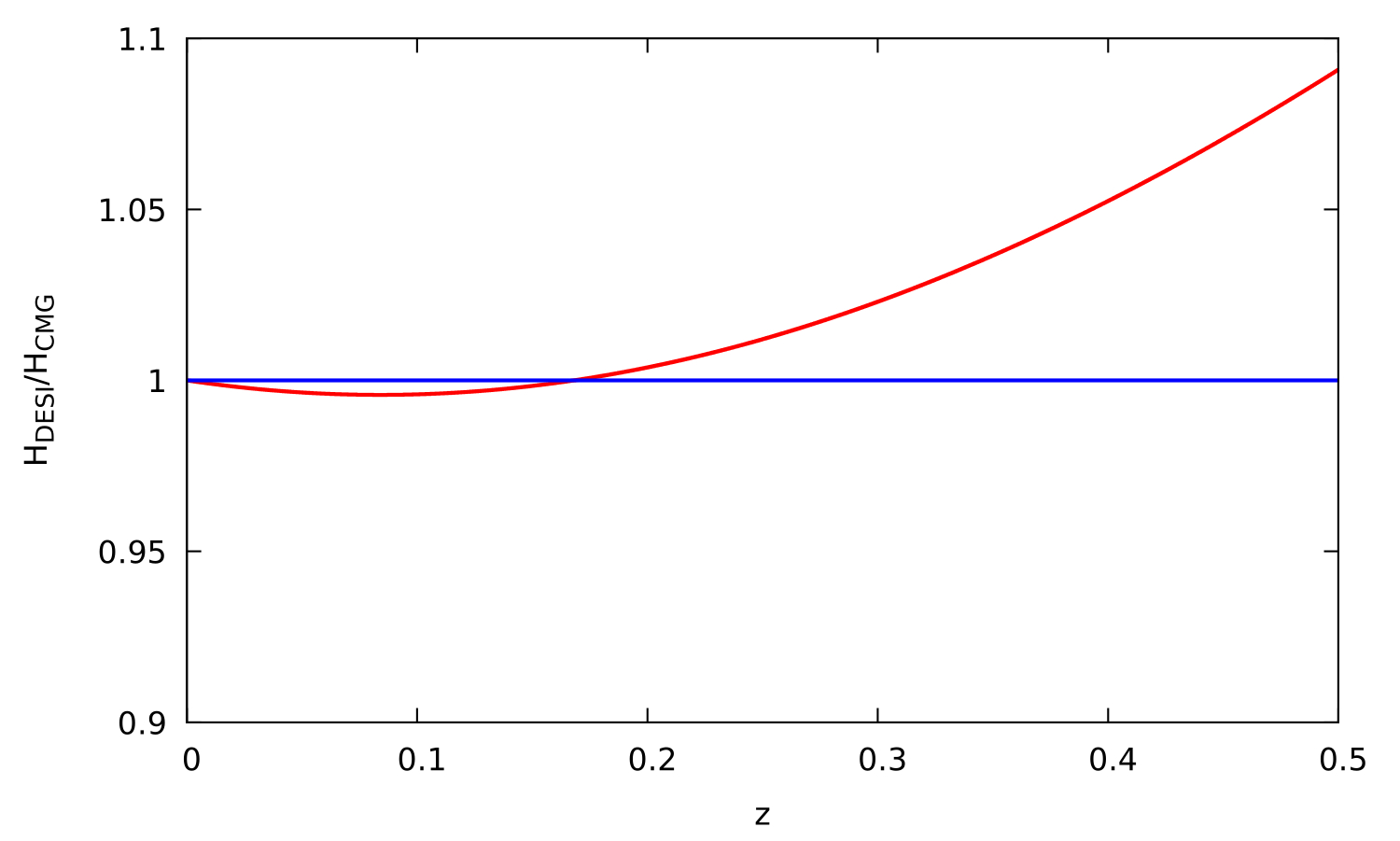}
\caption{Redshift evolution of the ratio of the Hubble scalar for a standard $\Lambda$CDM solution using recent DESI parameters and
  the Hubble scalar for the exact cosmological solution presented. For $z<0.2$ both solutions lie within half a percent of each other,
  showing the cosmological solution resulting from solving the empirically constrained galactic space-time field equations under a FLRW
metric to be an accurate representation of the present accelerated expansion of the Universe.}
\end{figure}

Comparison of the evolution of the Hubble scalar with redshift for the solution presented, to a standard $\Lambda$CDM
DESI+CMD+PantheonPlus (\cite{DESI2025}) $\Omega_{DE}=0.695$, $\Omega_{m}=0.305$, $\omega_{DE}=-0.8-0.7 z/(z+1)$ model show
both models to lie within 10\% of each other out to a redshift of 0.5, and within 0.5\% of each other out to a redshift of 0.2,
as shown in Figure (2). Thus, the present-day cosmology which results from taking the field equations obtained from the empirical
galactic space-time and applying them in a FLRW context closely matches observational inferences for the accelerated expansion of
the Universe. Just as the spherically symmetric and static solution presented for galactic space-times is only relevant in the low
density flat rotation curve regime, the cosmological solution shown will only be relevant in the present-time low density,
"dark energy" dominated regime.


\section{Discussion}
\label{sec:conclusions}

We have presented a simple minimally coupled action using a power law model for both the Ricci
scalar and the matter Lagrangian, and shown that the power law indexes and coupling constant
can be adjusted to fully reproduce a covariant description of galactic space-times in the
low acceleration regime, to leading order in $1/c^{2}$ under a spherically symmetric and static
approximation.

An accelerated cosmological model results which leads to a late time accelerated expansion without
the explicit inclusion of any dark energy component as such. The effective value of the equation of
state of the "dark energy" component which would be inferred under a GR interpretation of the modified
Friedmann equation resulting from a matter component, would be $\omega=-5/6$, compatible with recent
DESI observations.

In vacuum our theory is effectively given by 
\( f(R,\mathcal{L}_\mathcal{M}) = F(R) \propto R^{3/2} \).
As proven in e.g.  \cite{maldonado2026gravitational}, the condition for luminal gravitational waves in \(F(R) \) gravity  is 
$F'(R_\text{bck})/F''(R_\text{bck}) > 0 $.  Since in our case 
$F'(R_\text{bck})/F''(R_\text{bck}) = 2 R_\text{bck}$, which is indeed positive for both the static spherical symmetric  and the FLRW cases presented above,  gravitational waves for the presented action propagate at the speed of light.

Given the indices obtained, it is easy to see that if the standard Einstein-Hilbert terms are
added to the proposed action, they will dominate in the high density, strong gravity regime,
while the modified terms will dominate in the low density, low potential one.

Studying the rate of convergence of gravitational physics between the high and low density
regimes is necessary to constrain a complete gravitational theory. Future work should focus
on extending the results presented here to rotating systems, cosmological perturbations, and
non-vacuum solutions.

%
%
%
%
%

\begin{acknowledgments}
{\it Acknowledgements}—This work was supported by PAPIIT DGAPA-UNAM grants IN118325 and
IN102624.  XH, SM and SS acknowledge support from SECIHTI
(25006,26344,1228643).
\end{acknowledgments}

\bibliography{references}

\end{document}